\documentclass[12pt,draftcls,onecolumn]{IEEEtran}
\usepackage{amsmath}
\usepackage{amssymb}
\usepackage{cite}
\usepackage{color}
\usepackage{epsfig}
\usepackage{epsf}
\usepackage{rotating}
\usepackage{mathrsfs}
\usepackage{epsfig}
\usepackage{graphics}
\usepackage{theorem}

\newtheorem{thm}{Theorem}
\newtheorem{lemma}{Lemma}
\newtheorem{corr}{Corrolary}

\begin{document}
\begin{center}
\vskip 0.6cm
{\large \bf Asymptotic Analysis of Amplify and
 Forward Relaying\\ in a Parallel MIMO Relay Network} \\

\vskip 0.8cm Shahab Oveis Gharan, Alireza Bayesteh, and Amir K.
Khandani

\vskip 0.5cm

{\small Coding \& Signal Transmission Laboratory\\
Department of Electrical \& Computer Engineering\\
University of Waterloo\\
Waterloo, Ontario, Canada, N2L 3G1\\
}

\end{center}

\begin{abstract}
This paper considers the setup of a parallel MIMO relay network in
which $K$ relays, each equipped with $N$ antennas, assist the
transmitter and the receiver, each equipped with $M$ antennas, in
the half-duplex mode, under the assumption that $N\geq{M}$. This
setup has been studied in the literature like in \cite{nabar},
\cite{nabar2}, and \cite{qr}. In this paper, a simple scheme, the
so-called Incremental Cooperative Beamforming, is introduced and
shown to achieve the capacity of the network in the asymptotic case
of $K\rightarrow{\infty}$ with a gap no more than
$O\left(\frac{1}{\log\left(K\right)}\right)$. This result is shown
to hold, as long as the power of the relays scales as
$\omega\left(\frac{\log^9\left(K\right)}{K}\right)$. Finally, the
asymptotic SNR behavior is studied and it is proved that the
proposed scheme achieves the full multiplexing gain, regardless of
the number of relays.
\end{abstract}

\section{Introduction}

\subsection{Motivation}

In recent years, Multiple-input Multiple-output (MIMO) wireless
systems have received significant attention. It has been shown that
MIMO wireless systems have the ability to
 simultaneously enhance the multiplexing gain (degrees of freedom) and the diversity (reliability) of the Rayleigh fading channel \cite{telatar},\cite{foschini},\cite{
zheng_tse}. The relay channel, which was first introduced by Van-der
Meulen  in 1971 \cite{ van_der_meulen}, has been reconsidered in
recent years to improve the coverage, reliability, and reduce the
interference in the multi-user wireless networks. The main idea is
to employ some extra nodes in the network to aid the
transmitter/receiver in sending/receiving the signal to/from the
other end. In this way, the supplementary nodes act as (spatially)
distributed antennas assisting the signal transmission and
reception.

After some recent information-theoretic results on the MIMO
point-to-point Rayleigh fading channels
\cite{telatar},\cite{foschini},\cite{ zheng_tse}, there has been
growing interest in studying the impact of MIMO systems in more
complex wireless networks. Some promising results have been
published on MIMO Multiple-Access and Broadcast channels in
\cite{tse_mac},\cite{yu},\cite{goldsmith},\cite{shamai}, and
\cite{tse_mac_tradeoff}. However, there are still only a few results
known concerning the MIMO relay networks. Moreover, no
capacity-achieving strategy is known for the Gaussian relay channel.

This paper analyzes the performance of a parallel MIMO relay
network. Our focus is on the Amplify and Forward (AF) strategy. Not
only the AF strategy offers low complexity and delay, but also it
performs well in our setup.

\subsection{History}

The classical relay channel was first introduced by Van-der Meulen
in 1971 \cite{ van_der_meulen}. In \cite{van_der_meulen}, a node
defined as the relay enhances the transmission of information
between the transmitter and the receiver. The most important
relevant results have been published by Cover and El Gamal
\cite{cover}. In \cite{cover}, two different coding strategies are
introduced. In the first strategy, originally named ``cooperation",
and later known as ``decode-and-forward" (DF), the relay decodes the
transmitted message and cooperates with the transmitter to send the
message in the next block. In the second strategy, known as
``compress-and-forward" (CF), the relay compresses the received
signal and sends it to the receiver. The performance of the DF
strategy is limited by the quality of the transmitter-to-relay
channel, while CF's performance is mostly restricted by the quality
of the relay-to-receiver channel \cite{cover}. The drawback of using
CF strategy is that it employs no cooperation between the
transmitter and the relay at the receiver side. Hence, the CF
strategy is unable to exploit the power boosting advantage due to
the coherent addition of the signal of the transmitter and the
relay\cite{cover}.

More recently, several extensions of the relay channel have been
considered, e.g. in \cite{gallager, gastpar, gastpar2, xie}. Some of
these extensions consider a multiple-relay scenario in which several
nodes relay the message. The parallel relay channel is a special
case of the multiple relay channel in which the relays transmit
their data directly to the receiver. Besides studying the well-known
``compress-and-forward" and ``decode-and-forward" strategies, the
authors in \cite{gallager, gastpar} have also studied the
``amplify-and-forward" strategy where the relays simply amplify and
transmit their received data to the receiver. Despite its
simplicity, the AF strategy achieves a good performance. In fact,
\cite{gallager} shows that AF outperforms other strategies in many
scenarios. Moreover, \cite{gastpar} proves that AF achieves the
capacity of the Gaussian (single antenna) parallel relay network as
the number of relays increases.

References \cite{nabar, nabar2} extend the work of \cite{gastpar} to
the MIMO Rayleigh fading parallel relay network. Unlike the single
antenna parallel relay scenario, in this case the AF multipliers are
matrices rather than scalars. Hence, finding the optimum AF matrices
becomes challenging. Reference \cite{nabar} has proposed a coherent
AF scheme, called ``matched filtering", and proves that this scheme
follows the capacity of the channel with a constant gap in terms of
the number of relays in the asymptotic case of
$K\rightarrow{\infty}$. They also show that the achievable rate of
AF in parallel MIMO relay network grows linearly with the number of
antennas (reflecting the multiplexing gain) and grows
logarithmically in terms of the number of relays (reflecting the
distributed array gain \cite{nabar}).

Reference \cite{qr} presents a new AF scheme using the QR
decomposition of the forward and backward channels in each relay
that outperforms the other AF schemes for practical number of
relays.

\subsection{Contributions and Relation to Previous Works}

In this paper, we consider the AF strategy in the parallel MIMO
relay network. The channel is assumed to be Rayleigh fading and the
communication takes place in the half-duplex mode (i.e. the relays
can not transmit and receive simultaneously). We propose a new AF
protocol called ``Cooperative Beamforming Scheme" (CBS). Considering
the uplink channel (from the transmitter to the relays) as a
point-to-point channel, in CBS the relays cooperatively multiply the
channel matrix with its left eigenvector matrix. Hence, the relays
act like the spatially distributed antennas at the equivalent
receiver. The interesting point is that to perform such an
operation, each relay only needs to know its corresponding
sub-matrix of the beamforming matrix. For the outputs to be
coherently added at the receiver end, each relay has to apply zero
forcing beamforming to its corresponding downlink
 channel (the channel from each relay to the receiver). Here, the interesting result is that the overall
channel from the transmitter to the receiver becomes diagonal and
the overall Gaussian noise has independent components.

We show that the proposed scheme is optimum in the case of having
negligible noise in the downlink channel. However, the downlink
noise would degrade the system performance when one of the relays'
downlink channels is ill-conditioned. To enhance the performance of
CBS in general scenarios, this work introduces a variant of CBS
called ``Incremental Cooperative Beamforming Scheme" (ICBS). In
ICBS, the relays with ill-conditioned downlink channels are turned
off. This strategy improves the overall point-to-point channel from
the transmitter to the receiver. However, an interference term due
to turning some of the relays off will be included in the equivalent
point-to-point channel.

It is shown that for asymptotically large number of relays, one can
simultaneously mitigate the downlink noise and the interference term
due to the turned-off relays. As a result, the achievable rate of
ICBS converges to the capacity of parallel MIMO relay network with a
gap which scales as $O\left(\frac{1}{\log\left(K\right)}\right)$.
This result is stronger than the result of \cite{nabar} and
\cite{nabar2} in which they show that their scheme can
asymptotically ($K \to \infty$) achieve the capacity up to $O(1)$.
Also, our numerical results show that the achievable rate of ICBS
converges rapidly to the capacity, even for moderate number of
relays. Our results also demonstrate that the achievable rate of
ICBS, the maximum achievable rate of amplify and forward strategy,
the capacity of the parallel MIMO relay network, and the
point-to-point capacity of the uplink channel converge to each other
for asymptotically large number of relays.

We also show that the same result can be achieved by ICBS, as long
as the power of the relays scales as
$\omega\left(\frac{P}{K}\log^9\left(K\right)\right)$ \footnote{ $f
(n) = \omega (g (n))$ is equivalent to $\lim_{n \to \infty}
\frac{f(n)}{g(n)} =\infty$}. Finally, by analyzing the asymptotic
SNR behavior of the proposed scheme, it is proved that, unlike the
matched filtering scheme of B�cskei-Nabar-Oyman-Paulraj (BNOP)
which results in a zero multiplexing gain, our proposed scheme
achieves the full multiplexing gain, regardless of the number of
relays.

The rest of the paper is organized as follows. In section II, the
system model is introduced. In section III, the proposed AF scheme
is described. Section IV is dedicated to the asymptotic analysis of
the proposed scheme. Simulation results are presented in section V.
Finally, section VI concludes the paper.

\subsection{Notation}

Throughout the paper, the superscripts $^T$,$^H$ and $^*$ stand for
matrix operations of transposition, conjugate transposition, and
element-wise conjugation, respectively. Capital bold letters
represent matrices, while lowercase bold letters and regular letters
represent vectors and scalars, respectively. $\|\mathbf{v}\|$
denotes the norm of the vector $\mathbf{v}$ while $\|\mathbf{A}\|$
represents the frobenius norm of the matrix $\mathbf{A}$.
$|\mathbf{A}|$ denotes the determinant of the matrix $\mathbf{A}$
while $\|\mathbf{A}\|_{\star}$ represents the maximum absolute value
among the entries of $\mathbf{A}$.
 The notation
$\mathbf{A}^{\dag}$ stands for the pseudo inverse of the matrix
$\mathbf{A}$. The notation $\mathbf{A}\preccurlyeq\mathbf{B}$ is
equivalent to $\mathbf{B}-\mathbf{A}$ is a positive semi-definite
matrix. For any functions $f(n)$ and $g(n)$, $f(n)=O(g(n))$ is
equivalent to $\lim_{n \rightarrow \infty} \left| \frac{f(n)}{g(n)}
\right| < \infty$, $f(n)=o(g(n))$ is equivalent to $\lim_{n
\rightarrow \infty} \left| \frac{f(n)}{g(n)} \right| =0$,
$f(n)=\Omega(g(n))$ is equivalent to $\lim_{n \rightarrow \infty}
\frac{f(n)}{g(n)}
>0$, $f(n)\gtrsim g(n)$ is equivalent to $\lim_{n
\rightarrow \infty} \frac{f(n)}{g(n)} \geq 1$, $f(n)=\omega(g(n))$
is equivalent to $\lim_{n \rightarrow \infty}  \frac{f(n)}{g(n)}  =
\infty$, $f(n) \sim g(n)$ is equivalent to $\lim_{n \rightarrow
\infty} \frac{f(n)}{g(n)} =1$ and $f(n)=\Theta(g(n))$ is equivalent
to $\lim_{n \rightarrow \infty} \frac{f(n)}{g(n)} =c$, where
$0<c<\infty$.

\section{System Model}

\begin{figure}[hbt]
  \centering
  \includegraphics[scale=.6]{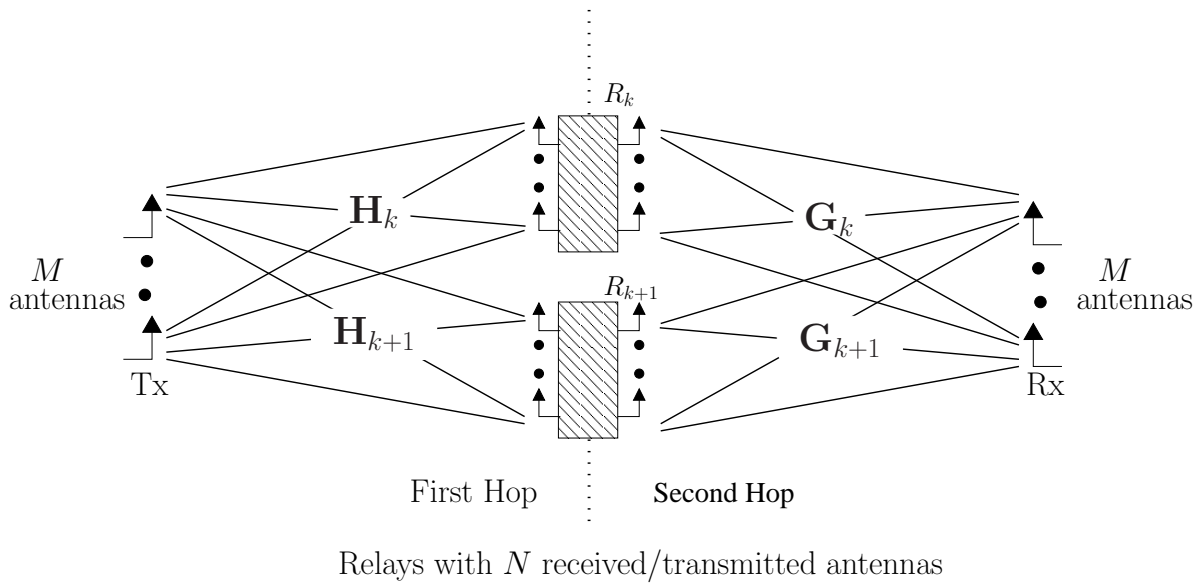}
\caption{A schematics of a parallel MIMO half-duplexing relay
network} \label{fig:system}
\end{figure}

The system model, as in \cite{nabar}, \cite{nabar2}, and \cite{qr},
is a parallel MIMO relay network with two-hop relaying and
half-dulplexing between the uplink and downlink channels. In other
words, the data transmission is performed in two time slots; in the
first time slot, the signal is transmitted from the transmitter to
the relays, and in the second time slot, the relays transmit data to
the receiver. Note that there is no direct link between the
transmitter and the receiver in this model.  The transmitter and the
receiver are equipped with $M$ antennas and each of the relays is
equipped with $N$ antennas. Throughout the paper, we assume that
$N\geq{M}$. The channel between the transmitter and the relays and
the channel between the relays and the receiver are assumed to be
frequency flat block Rayleigh fading. The channel from the
transmitter to the $k$th relay, $1\leq{k}\leq{K}$, is modeled as
\begin{equation}
\mathbf{r}_k=\mathbf{H}_k\mathbf{x}+\mathbf{n}_k,
\end{equation}
and the downlink channel is modeled as
\begin{equation}
\mathbf{y}=\sum_{k=1}^K{\mathbf{G}_k\mathbf{t}_k}+\mathbf{z},
\end{equation}
where the channel matrices $\mathbf{H}_k$ and $\mathbf{G}_k$ are
i.i.d. complex Gaussian matrices with zero mean and unit variance.
$\mathbf{n}_k\sim{\mathcal{CN}(\mathbf{0}, \mathbf{I}_N)}$ and
$\mathbf{z}\sim{\mathcal{CN}(\mathbf{0}, \mathbf{I}_M)}$ are
Additive White Gaussian Noise (AWGN) vectors, $\mathbf{r}_k$ and
$\mathbf{t}_k$ are the $k$th relay's received and transmitted
signal, respectively, and $\mathbf{x}$ and $\mathbf{y}$ are the
transmitter's and the receiver's signal, respectively.
 $\mathbf{H}_k$ and $\mathbf{G}_k$ are of the sizes $N{\times}M$ and $M{\times}N$, respectively (figure \ref{fig:system}).

The task of amplify and forward (AF) relaying is to find the matrix
$\mathbf{F}_k$ for each relay to be multiplied by its received
signal to produce the relay's output as
$\mathbf{t}_{k}=\mathbf{F}_{k}\mathbf{r}_{k}$. In this way, the
entire source-destination channel is modeled as
\begin{equation}
\mathbf{y}=\left(\sum_{k=1}^{K}{\mathbf{G}_k\mathbf{F}_k\mathbf{H}_k}\right)\mathbf{x}
+\sum_{k=1}^{K}{\mathbf{G}_k\mathbf{F}_k\mathbf{n}_k}+\mathbf{z}.
\end{equation}

In addition, the power constraints
$\mathbb{E}[\mathbf{x}^H\mathbf{x}]\leq{P_s}$ and
$\mathbb{E}_{\mathbf{x},
\mathbf{n}_k}[\mathbf{t}_k^H\mathbf{t}_k]\leq{P_r}$ must be
satisfied for the transmitted signals of the transmitter and the
relays, respectively. We assume $P_r=P_s=P$ throughout the paper,
except in Theorem
 \ref{thm:second_uneq_power}, where
we study the case $P_r<P_s=P$.

\section{Proposed Method}\label{section:proposed_method}

\subsection{Cooperative Beamforming Scheme}

\begin{figure}[hbt]
  \centering
  \includegraphics[scale=.5]{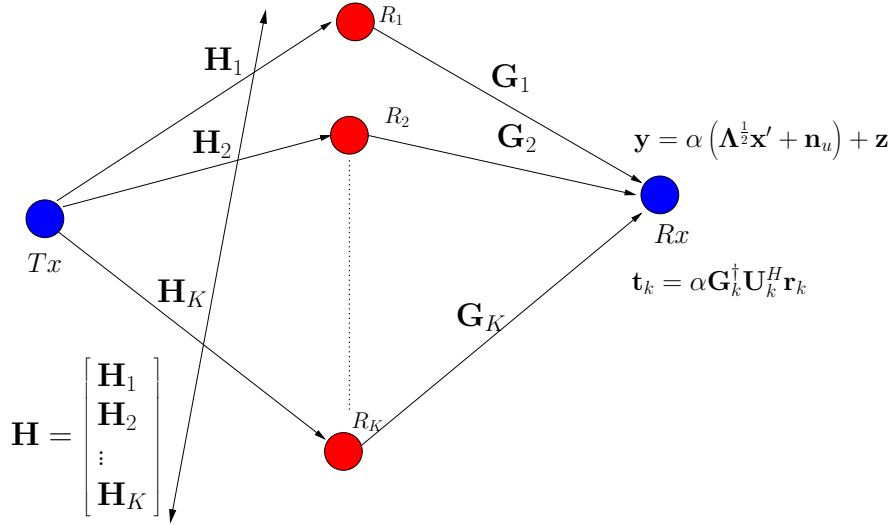}
\caption{Cooperative Beamforming Scheme}
\label{fig:cbs}
\end{figure}

The equivalent uplink channel can be represented as
$\mathbf{H}^T=\left[\mathbf{H}_1^T|\mathbf{H}_2^T|{\cdots}|\mathbf{H}_K^T\right]^T$.
 By applying Singular Value Decomposition (SVD) to $\mathbf{H}$, we
have
$\mathbf{H}=\mathbf{U}\mathbf{\Lambda}^{\frac{1}{2}}\mathbf{V}^H$.
Therefore, the diagonal matrix $\mathbf{\Lambda}$ has at most $M$
nonzero diagonal entries corresponding to the nonzero singular
values of $\mathbf{H}$. Consequently, we can rearrange the SVD such
that $\mathbf{U}$ is of size $NK\times{M}$ while $\mathbf{V}$ and
$\mathbf{\Lambda}$ are $M\times{M}$ matrices. $\mathbf{U}$ can be
partitioned to $M\times{N}$ sub-matrices as
$\mathbf{U}=\left[\mathbf{U}_1^T|\mathbf{U}_2^T|{\cdots}|\mathbf{U}_K^T\right]^T.$
 Suppose the $k$th relay multiplies its received signal by
$\mathbf{U}_k^H$, then passes it through the zero-forcing matrix
$\mathbf{G}_k^{\dagger}$, and finally amplifies it with a constant
scalar $\alpha$ independent of $k$; equivalently, we have
$\mathbf{F}_k=\alpha\mathbf{G}_k^{\dagger}\mathbf{U}_k^H$. At the
receiver side, we have (figure \ref{fig:cbs})
\begin{eqnarray}
\mathbf{y}&=&\alpha\sum_{k=1}^{K}{\mathbf{G}_k\mathbf{t}_k}+\mathbf{z}\nonumber\\
&=&\alpha\sum_{k=1}^{K}{\mathbf{G}_k\mathbf{G}_k^{\dagger}\mathbf{U}_k^H\mathbf{r}_k}+\mathbf{z}\nonumber\\
&=&\alpha{\mathbf{U}^H\mathbf{r}}+\mathbf{z}\nonumber\\
&=&\alpha\mathbf{U}^H\left(\mathbf{H}\mathbf{x}+\mathbf{n}\right)+\mathbf{z}\nonumber\\
&=&\alpha\left(\mathbf{\Lambda}^{\frac{1}{2}}\mathbf{V}^H\mathbf{x}+\mathbf{n}_u\right)+\mathbf{z},
\end{eqnarray}
where
$\mathbf{n}=\left[\mathbf{n}_1^T|\mathbf{n}_2^T|{\cdots}|\mathbf{n}_K^T\right]^T$
,
$\mathbf{r}=\left[\mathbf{r}_1^T|\mathbf{r}_2^T|{\cdots}|\mathbf{r}_K^T\right]^T$,
and
$\mathbf{n}_u=\mathbf{U}^H\mathbf{n}\sim{\mathcal{CN}(\mathbf{0},
\mathbf{I}_M)}$. If the transmitter beamforms its data vector as
$\mathbf{x}=\mathbf{V}\mathbf{x}^\prime$, the end-to-end channel
becomes
\begin{equation}\label{eq:trans_rec_pro}
\mathbf{y}=\alpha\left(\mathbf{\Lambda}^{\frac{1}{2}}\mathbf{x}^{\prime}+\mathbf{n}_u\right)+\mathbf{z}.
\end{equation}

Equation (\ref{eq:trans_rec_pro}) shows that the end-to-end channel
is diagonal and the noise vector is white Gaussian. Note that the
complexity of the decoder in such a channel is linear in terms of
the number of transmitter's antennas, $M$, and also there is no
interference among different data streams. In fact, the output
signals of the relays not only do not interfere with each other, but
also add constructively at the receiver side. Moreover, as it is
shown in section IV, for $\alpha\rightarrow\infty$, the achievable
rate of such a scheme converges to the point-to-point capacity of
the uplink channel which is shown to be an upper-bound on the
capacity of the parallel relay system.

The problem is that the value of $\alpha$ is dominated by
\begin{equation}\label{eq:alpha}
\alpha=\sqrt{\frac{P}{\max_{k}{\mathbb{E}_{\mathbf{x}, \mathbf{n}_k}
\left[\left\|\mathbf{G}_k^{\dagger}\mathbf{U}_k^H\mathbf{r}_k\right\|^2
\right]}}}.
\end{equation}

This guarantees that the output power of all relays is less than or
equal to $P$. However, by applying (\ref{eq:alpha}), the value of
$\alpha$ could be small in the cases where the downlink channel of
any of the relays is ill conditioned. This means that while the
output power of the worst relay (according to (\ref{eq:alpha})) is
equal to the maximum possible value, i.e. $P$, there may be many
relays with the output power far less than $P$. This phenomenon
degrades the performance, as in this case the downlink noise,
$\mathbf{z}$, would be the dominant noise in
(\ref{eq:trans_rec_pro}).

\subsection{Incremental Cooperative Beamforming Scheme (ICBS)}

\begin{figure}[hbt]
  \centering
  \includegraphics[scale=.5]{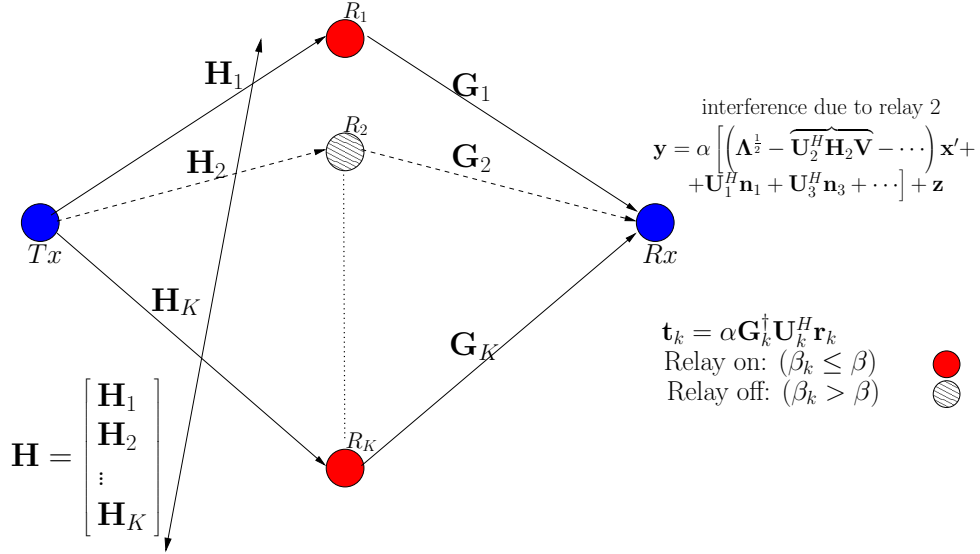}
\caption{Incremental Cooperative Beamforming Scheme}
\label{fig:icbs}
\end{figure}

As the number of relays increases, we expect (as shown in
(\ref{eq:alpha})) to have smaller values of $\alpha$ with high
probability. In other words, there is a higher chance of having at
least one ill-conditioned downlink channel among the relays. In this
case, we can select a subset of relays which are in good condition
and turn off the rest. In this variant of CBS, we select a subset of
relays which results in a high value of $\alpha$. Defining
$\beta_{k} \triangleq \mathbb{E}_{\mathbf{x}, \mathbf{n}_k}
\left[\left\|\mathbf{G}_k^{\dagger}\mathbf{U}_k^H\mathbf{r}_k\right\|^2
\right]$, we activate the relays which satisfy $\beta_k\leq{\beta}$,
where $\beta$ is a predefined threshold. In this manner, it is
guaranteed that $\alpha\geq{\sqrt\frac{P}{\beta}}$. This improvement
in the value of $\alpha$ is realized at the expense of turning off
some of the relays, creating interference in the equivalent
point-to-point channel. More precisely, by defining
$\mathcal{A}=\left\{k|\beta_k>\beta\right\}$, we have (figure
\ref{fig:icbs})
\begin{equation}\label{eq:interfer}
\mathbf{y}=\alpha\left(\left(\mathbf{\Lambda}^{\frac{1}{2}}
-\sum_{k\in\mathcal{A}}{\mathbf{U}_k^H\mathbf{H}_k\mathbf{V}}\right)\mathbf{x}^\prime
+\sum_{k\in\mathcal{A}^c}{\mathbf{U}_k^H\mathbf{n}_k}\right)+\mathbf{z}.
\end{equation}
As (\ref{eq:interfer}) shows, by decreasing the value of $\beta$,
one can guarantee a large value of $\alpha$ while increasing the gap
of the equivalent channel matrix to
$\mathbf{\Lambda}^{\frac{1}{2}}$. It will be shown in the next
section that for large number of relays, it is possible to guarantee
both having a large value of $\alpha$ and a small deviation from
$\mathbf{\Lambda}^{\frac{1}{2}}$. Moreover, we show that by
appropriately choosing the value of $\beta$, the rate of such a
scheme would be at most $O\left(\frac{1}{\log\left(K\right)}\right)$
below the corresponding capacity.

\subsection{A Note on CSI Assumption}
In the BNOP scheme, it is assumed that each relay knows its
corresponding forward and backward channels, i.e. $\mathbf{H}_k$ and
$\mathbf{G}_k$, and at the receiver side, the effective signal power
and the effective interference plus noise power are known for each
antenna. However, in CBS and ICBS, it is assumed that the
transmitter knows the uplink channel, i.e.
$\mathbf{H}_1,\cdots,\mathbf{H}_K$, and sends the $N \times M$
matrix $\mathbf{U}_k$ to the $k$'th relay, $k=1,\cdots,K$. This
assumption is reasonable when the uplink channel is slow-fading; for
example, in the case that the transmitter and all the relay nodes
are fixed. Furthermore, similar to the BNOP scheme, we assume that
each relay knows its forward channel, i.e. $\mathbf{G}_k$. In
addition, in CBS, it is assumed that the value of $\alpha$ is set by
negotiating between the relays through sending their corresponding
$\beta_k$ to the transmitter. This assumption is not required in
ICBS, as the value of $\alpha$ can be set as
$\alpha=\sqrt{\frac{P}{\beta}}$, where $\beta$ is a predefined
threshold. Finally, in both CBS and ICBS, it is assumed that the
receiver has the perfect knowledge about the equivalent
point-to-point channel from the transmitter to the receiver. This
information can be obtained through sending pilot signals by the
transmitter, amplified and forwarded at the relay nodes in the same
manner  as the information signal. In CBS, as the equivalent
point-to-point channel is diagonal, this assumption is equivalent to
knowing the equivalent signal to noise ratio at each antenna.

\section{Asymptotic Analysis}{\label{sec:Per_Anal}}

In this section, we consider the asymptotic behavior
($K\rightarrow\infty$) of the achievable rate of ICBS. We show that
by properly choosing the value of $\beta$, the achievable rate of
ICBS converges rapidly to the capacity (the difference approaches
zero as $O\left(\frac{1}{\log(K)}\right)$). The sequence of proof is
as follows. In Lemma 1, we relate $\mathbb{P}\left[v>\xi\right]$
(the probability that the norm of interference term defined in
equation (\ref{eq:interfer}) exceeds a certain threshold) to
${\mathbb{P}}[k\in\mathcal{A}]$ (the probability of turning off a
relay) and $\mathbb{P}[\|\mathbf{U}_k\|^2>\gamma]$ (the probability
of having a sub-matrix with a large norm in the unitary matrix
obtained from the SVD of $\mathbf{H}$). In Lemma 2, we bound
$\mathbb{P}[\|\mathbf{U}_k\|^2>\gamma]$. In Lemma 3, we bound
${\mathbb{P}}[k\in\mathcal{A}]$. As a result, in Lemma 4, we show
that by properly choosing the value of $\beta$, with high
probability, one can simultaneously reduce the effect of the
interference to $o(K)$ and maintain a large value of $\alpha$. In
Lemma 5, we show that with high probability, the minimum singular
value of $\mathbf{H}$ scales as $O(K)$. Putting Lemmas 4 and 5
together, with high probability, the ratio of the power of
interference to the power of signal approaches zero. Finally, in
Theorem 1, we prove the main result by showing that the achievable
rate of ICBS converges to the capacity of the uplink channel. This
is proved using the fact that the capacity of the uplink channel is
an upper-bound on the capacity of parallel MIMO relay network. As a
consequence stated in corollary 1, the achievable rate of ICBS, the
achievable rate of the AF protocol, the point-to-point capacity of
the uplink channel, and the capacity of the parallel MIMO relay
network are asymptotically equal. As another consequence, the
difference of the rates scales as $O(\frac{1}{\log(K)})$.

Using the proof of Lemma 4 and Theorem 1, Theorem 2 shows that as
long as the power of relays behaves as
$P_r(K)=\omega\left(\frac{P}{K}\log^9\left(K\right)\right)$, the
same rate is achievable by ICBS. Finally, in Theorem 3, we study the
asymptotic SNR behavior of CBS and ICBS, and show that, unlike the
matched filtering scheme of BNOP, CBS and its variant achieve the
full multiplexing gain, regardless of the number of relays.

\begin{lemma}
Consider a parallel MIMO relay network with $K$ relays using ICBS.
We have
\begin{eqnarray}\label{eq:t11}
\mathbb{P}\left[v>\xi\right]&\leq&\frac{MNK^2}{\xi}\left(
\mathbb{P}[B_k]+\gamma{\mathbb{P}}[A_k]\right),
\end{eqnarray}
where $v$ is defined as
$v=\left\|\sum_{k\in\mathcal{A}}{\mathbf{U}_k^H\mathbf{H}_k}\right\|^2$
, and $A_k$ and $B_k$ are indicator variables defined as
$A_k\equiv{(k\in\mathcal{A})}$ and
$B_k\equiv{(\|\mathbf{U}_k\|^2>\gamma)}$, respectively.
\end{lemma}
\begin{proof}
Let us define
$\mathbf{U}_{\mathcal{A}}=\left[\mathbf{U}_k^T|k\in\mathcal{A}\right]^T$
and
$\mathbf{H}_{\mathcal{A}}=\left[\mathbf{H}_k^T|k\in\mathcal{A}\right]^T$.
We have
\begin{eqnarray}
\mathbb{P}\left[v>\xi\right]&=&\mathbb{P}\left[\|\mathbf{U}_{\mathcal{A}}^H\mathbf{H}_{\mathcal{A}}\|^2>\xi\right]\nonumber\\
&\stackrel{(a)}{\leq}&\frac{\mathbb{E}\left[\|\mathbf{U}_{\mathcal{A}}^H\mathbf{H}_{\mathcal{A}}\|^2\right]}{\xi}\nonumber\\
&\stackrel{(b)}{\leq}&\frac{\mathbb{E}\left[\|\mathbf{U}_{\mathcal{A}}\|^2\|\mathbf{H}_{\mathcal{A}}\|^2\right]}{\xi}\nonumber\\
&\stackrel{(c)}{\leq}& \frac{\mathbb{E}\left[\|\mathbf{U}_{\mathcal{A}}\|^2\|\mathbf{H}\|^2\right]}{\xi}\nonumber\\
&\stackrel{(d)}{=}&\frac{\mathbb{E}\left[\|\mathbf{U}_{\mathcal{A}}\|^2\right]\mathbb{E}\left[\|\mathbf{H}\|^2\right]}{\xi}\nonumber\\
&=&\frac{MNK\mathbb{E}\left[\|\mathbf{U}_{\mathcal{A}}\|^2\right]}{\xi}\label{eq:t12}
\end{eqnarray}
Here, Markov inequality is applied to derive inequality $(a)$. $(b)$
is obtained by applying the norm product inequality on
matrices\footnote{Assuming $\mathbf{A}$ and $\mathbf{B}$ two
matrices of sizes $m\times{n}$ and $n\times{k}$, correspondingly, we
have
$\|\mathbf{A}\mathbf{B}\|^2\leq\|\mathbf{A}\|^2\|\mathbf{B}\|^2$
\cite{matrix_book} .}. $(c)$ results from the fact that
$\|\mathbf{H}_{\mathcal{A}}\|^2 \leq \|\mathbf{H}\|^2$. Finally,
equation $(d)$ follows from the fact that the left unitary matrix,
i.e. $\mathbf{U}$, resulted from the SVD  of an i.i.d. complex
Gaussian matrix, is independent of its singular value matrix, i.e.
$\mathbf{\Lambda}^{\frac{1}{2}}$ ,\cite{hochwald}, and the fact that
$\|\mathbf{H}\|^2$ is a function of $\mathbf{\Lambda}$.

To upper-bound
$\mathbb{E}\left[\|\mathbf{U}_{\mathcal{A}}\|^2\right]$, we have
\begin{eqnarray}
\mathbb{E}\left[\|\mathbf{U}_{\mathcal{A}}\|^2\right]&=&\mathbb{E}\left[\sum_{k=1}^K{A_k\|\mathbf{U}_{k}\|^2}\right]\nonumber\\
&\stackrel{(a)}{=}&K\mathbb{E}\left[{A}_k\|\mathbf{U}_{k}\|^2\right]\nonumber\\
&=&K\mathbb{E}\left[\|\mathbf{U}_{k}\|^2|A_k\right]\mathbb{P}[A_k]\nonumber\\
&=&K\mathbb{E}\left[\|\mathbf{U}_{k}\|^2|A_k,B_k\right]\mathbb{P}[A_k,
B_k]\nonumber\\
&+&K\mathbb{E}\left[\|\mathbf{U}_{k}\|^2|A_k,B_k^c\right]\mathbb{P}[A_k,
B_k^c]\nonumber\\
&\stackrel{(b)}{\leq}&K\left(\mathbb{P}[A_k, B_k]+\gamma\mathbb{P}[A_k, B_k^c]\right)\nonumber\\
&\stackrel{(c)}{\leq}&K\left(\mathbb{P}[B_k]+\gamma\mathbb{P}[A_k]\right),\label{eq:t13}
\end{eqnarray}
where $(a)$ follows from the fact the channels are symmetric, $(b)$
follows from the fact that the norm of $\mathbf{U}_k$ is
upper-bounded by 1 and conditioned on the event $B_k^c$, it is
upper-bounded by $\gamma$, and finally $(c)$ follows from the basic
probability inequalities. Combining inequalities (\ref{eq:t12}) and
(\ref{eq:t13}) completes the proof.
\end{proof}
\begin{lemma}
Consider a $KN\times{M}$ Unitary matrix $\mathbf{U}$, where its
columns $\mathbf{U}_i$, $i=1,\cdots,M$, are isotropically
distributed unit vectors in $\mathbb{C}^{NK\times{1}}$. Let
$\mathbf{W}$ be an arbitrary $N\times{M}$ sub-matrix of
$\mathbf{U}$. Then, for a predefined value of $M$ and $N$ and
assuming $\gamma=\omega\left(\frac{1}{K}\right)$, as $K \to \infty$,
we have
\begin{equation}
\mathbb{P}\left[\|\mathbf{W}\|^2\geq\gamma\right]=O\left(
{{\left(K\gamma\right)^{(N-1)}e^{-\frac{\gamma}{M}NK}}}\right)
\end{equation}
\end{lemma}
\begin{proof}
See Appendix A.
\end{proof}
\begin{lemma}
For a small enough value of $\delta$, we have
\begin{equation}\label{eq:t30}
\mathbb{P}[A_k]\leq\mathbb{P}[B_k]+c_1\sqrt{\delta}+c_2e^{-\frac{d}{\sqrt{\delta}}},
\end{equation}
where $\delta=\frac{\gamma}{\beta}$, and $c_1, c_2$ and $d$ are positive constant parameters independent of $K, \beta$, and $\gamma$.
\end{lemma}
\begin{proof}
Assume $k$'th relay is off. Hence, we have
\begin{equation}\label{eq:t31}
\beta<\mathbb{E}_{\mathbf{x}, \mathbf{n}_k}
\left[\left\|\mathbf{G}_k^{\dagger}\mathbf{U}_k^H\mathbf{r}_k\right\|^2
\right]\stackrel{(a)}{\leq}
{\lambda_{\min}^{-1}(\mathbf{G}_k)}\|\mathbf{U}_k\|^2\left(1+P\|\mathbf{H}_k\|^2\right).
\end{equation}
Here, $(a)$ follows from the product norm inequality of matrices and
independency of the noise from other random variables in the system.
Defining the events
\begin{eqnarray}
C_k&\equiv&\left(\lambda_{\min}(\mathbf{G}_k)<\frac{\|\mathbf{U}_k\|^2}{\beta}(1+P\|\mathbf{H}_k\|^2)\right),\\
D_k&\equiv&\left(\lambda_{\min}(\mathbf{G}_k)<\delta\left(1+P\|\mathbf{H}_k\|^2\right)\right),
\end{eqnarray}
we have
\begin{eqnarray}\label{eq:t32}
\mathbb{P}[A_k]&\stackrel{(a)}{\leq}&\mathbb{P}[C_k]\nonumber\\
&\stackrel{}{=}&\mathbb{P}[C_k\cap{B_k}]+\mathbb{P}[C_k\cap{B}_k^c]\nonumber\\
&\stackrel{(b)}{\leq}&\mathbb{P}[{B_k}]+\mathbb{P}[C_k|{B}_k^c]\mathbb{P}[{B}_k^c]\nonumber\\
&\stackrel{(c)}{\leq}&\mathbb{P}[{B_k}]+\mathbb{P}[D_k|{{B}_k^c}]\mathbb{P}[{B}_k^c]\nonumber\\
&\stackrel{(d)}{\leq}&\mathbb{P}[{B_k}]+\mathbb{P}[D_k],
\end{eqnarray}
where $(a)$ results from (\ref{eq:t31}), $(b)$ and $(d)$ follow from
basic probability inequalities and $(c)$ follows from the fact that
conditioned on $\|\mathbf{U}_k\|^2\leq\gamma$, we have
$\frac{\|\mathbf{U}_k\|^2}{\beta}(1+P\|\mathbf{H}_k\|^2) <
\delta\left(1+P\|\mathbf{H}_k\|^2\right)$, which incurs that
$C_k\subseteq{D_k}$. Defining $\mathbf{W}_k$ as the submatrix
defined on the first $M$ rows of $\mathbf{G}_k$, we have
\begin{eqnarray}\label{eq:t33}
\mathbb{P}[D_k]&\leq&\mathbb{P}\left[\left(\lambda_{\min}(\mathbf{G}_k)\leq\sqrt{\delta}\right)\bigcup\left(1+P\|\mathbf{H}_k\|^2\geq\frac{1}{\sqrt{\delta}}\right)\right]\nonumber\\
&\stackrel{(a)}{\leq}&\mathbb{P}\left[\lambda_{\min}(\mathbf{G}_k)\leq\sqrt{\delta}\right]+\mathbb{P}\left[1+P\|\mathbf{H}_k\|^2\geq\frac{1}{\sqrt{\delta}}\right]\nonumber\\
&\stackrel{(b)}{\leq}&\mathbb{P}\left[\lambda_{\min}(\mathbf{W}_k)\leq\sqrt{\delta}\right]+\mathbb{P}\left[1+P\|\mathbf{H}_k\|^2\geq\frac{1}{\sqrt{\delta}}\right]\nonumber\\
&\stackrel{(c)}{=}&\int_{x=0}^{\sqrt{\delta}}{Me^{-Mx}dx}
+\frac{1}{\Gamma(MN)}
\int_{x=\frac{1}{P}\left(\frac{1}{\sqrt{\delta}}-1\right)}^{\infty}{x^{MN-1}e^{-x}dx}\nonumber\\
&\leq&M\sqrt{\delta}+\left[\sum_{m=0}^{MN-1}{\frac{x^me^{-x}}{m!}}\right]_{x=\frac{1}{P}\left(\frac{1}{\sqrt{\delta}}-1\right)}\nonumber\\
&\stackrel{(d)}{\leq}&M\sqrt{\delta}+MNe^{-\frac{1}{2P}\left(\frac{1}{\sqrt{\delta}}-1\right)}\nonumber\\
&=&M\sqrt{\delta}+MNe^{\frac{1}{2P}}e^{-\frac{1}{2P\sqrt{\delta}}}
\end{eqnarray}
Here, $(a)$ results from the union bound, $(b)$ results from the
fact that
$\lambda_{\min}\left(\mathbf{G}_k\right)\geq\lambda_{\min}\left(\mathbf{W}_k\right)$
which can be shown easily based on the definition of the singular
values of a matrix, $(c)$ results from applying the probability
density function of the minimum singular value of square i.i.d.
complex Gaussian matrix, derived in \cite{edelman}, and also the
fact that $\| \mathbf{H}_k\|^2$ has Chi-Square distribution with
$2MN$ degrees of freedom, and finally, $(d)$ results from the
assumption that $\delta$ is small enough such that $\forall{}m,
0\leq{m}<MN$, we have
 $\left(\frac{1}{P}(\frac{1}{\sqrt{\delta}}-1)\right)^m<e^{\frac{1}{2P}\left(\frac{1}{\sqrt{\delta}}-1\right)}$. By Combining the results of (\ref{eq:t32}) and
(\ref{eq:t33}), we obtain (\ref{eq:t30}) and this completes the
proof.
\end{proof}

Next, we apply Lemmas 1, 2, and 3 to prove that for large values of
$K$, by properly choosing the value of $\beta$, ICBS can
simultaneously achieve a large value of $\alpha$ and reduce the
interference to $o(K)$, with a high probability.
\begin{lemma}
By assigning $\beta=\frac{1}{\log(K)}$ and
$\gamma=\frac{2\log(K)}{K}$, ICBS simultaneously achieves
\begin{eqnarray}
\alpha=\Omega\left(\sqrt{\log(K)}\right),\label{eq:t41}\\
\mathbb{P}\left[v > \frac{K}{\log^2
(K)}\right]=O\left(\frac{\log^4\left(K\right)}{\sqrt{K}}\right)\label{eq:t42},
\end{eqnarray}
where $v$ is defined in Lemma 1.
\end{lemma}
\begin{proof}
Having $\beta=\frac{1}{\log(K)}$, the value of $\alpha$ would be
\begin{equation}
\alpha=\sqrt{\frac{P}{\max_{k\in\mathcal{A}^c}{\beta_k}}}\geq\sqrt{\frac{P}{\beta}}=\Omega\left(\sqrt{\log(K)}\right),
\end{equation}
and this results in (\ref{eq:t41}). Assuming
$\xi=\frac{K}{\log^2(K)}$, we have
\begin{eqnarray}\label{eq:t43}
\mathbb{P}\left[v>\xi\right]&\stackrel{(a)}{\leq}&MNK\log^2(K)\left(\mathbb{P}[B_k]+\frac{2 \log(K)}{K}\mathbb{P}[A_k]\right)\nonumber\\
&\stackrel{(b)}{\leq}&MNK\log^2(K)\left[\mathbb{P}[B_k]+ \frac{2\log(K)}{K}\left(\mathbb{P}[B_k]+c_1\sqrt{2 \frac{\log^2(K)}{K}}+{}\right.\right.\nonumber\\
&&\left.\left.{}c_2e^{-d\sqrt{\frac{K}{2 \log^2(K)}}}\right)\right]\label{eq: alireza1}\\
&\stackrel{(c)}{\leq}&2MNK\log^2(K)\mathbb{P}[B_k]+ 2MN\sqrt{2}c_1\frac{\log^{4}(K)}{\sqrt{K}}+2MNc_2\frac{\log^3(K)}{K}\nonumber\\
&=&2MNK\log^2(K)\mathbb{P}[B_k]+ O\left(\frac{\log^4(K)}{\sqrt{K}}\right)\nonumber\\
&\stackrel{(d)}{=}&2MNK\log^2(K)O\left(\left(\log(K)\right)^{(N-1)}e^{-\frac{2{N}}{M}\log(K)}\right) + O\left(\frac{\log^4(K)}{\sqrt{K}}\right)\nonumber\\
&\stackrel{(e)}{=}&O\left(\frac{\log^4(K)}{\sqrt{K}}\right).
\end{eqnarray}
Here, $(a)$ follows from Lemma 1, $(b)$ follows from Lemma 3, $(c)$
follows the assumption that $K$ is large enough such that $2
\log(K)<K$ and $d\sqrt{2}\frac{\sqrt{{K}}}{\log(K)}\geq{\log(K)}$ ,
$(d)$ follows from Lemma 2, and $(e)$ follows from the fact that
$\frac{2{N}}{M}\geq2$, which incurs that
$$K\log^2(K)O\left(\left(\log(K)\right)^{(N-1)}e^{-\frac{2{N}}{M}\log(K)}\right)
\sim O \left(\frac{\log^{N+1} (K)}{K}\right) \sim
o\left(\frac{\log^4(K)}{\sqrt{K}}\right).$$ This completes the proof
of Lemma 4.
\end{proof}
Although with the threshold value stated by Lemma 4, the
interference term  may tend to infinity in terms of $K$, the signal
term tends to infinity more rapidly. In fact, as the following Lemma
shows, the singular values of the whole uplink channel matrix behave
as $O(K)$ with probability 1, as $K \to \infty$.
\begin{lemma}
Let $\mathbf{A}$ be an $r\times{s}$ matrix whose entries are i.i.d
complex Gaussian random variables with zero mean and unit variance.
Assume that $r$ is fixed and $s$ tends to infinity. Then, with
probability one $\lambda_{\min} (\mathbf{A})\sim{s},$ or more
precisely,
\begin{equation}
\mathbb{P}\left[\lambda_{\min} (\mathbf{A})\sim{s
\left(1+O\left(\sqrt[4]{\frac{\log(s)}{s}}\right)\right)}\right]\gtrsim{}1-O\left(\frac{1}{s\sqrt{\log{(s)}}}\right),
\end{equation}
where $\lambda_{\min} (\mathbf{A})$ denotes the minimum singular
value of $\mathbf{A}\mathbf{A}^H$.
\end{lemma}
\begin{proof}
See Appendix B.
\end{proof}
Next, we prove the main theorem of this section.
\begin{thm}\label{thm:main_thm}
By setting the threshold as $\beta=\frac{1}{\log(K)}$, the
achievable rate of the proposed ICBS converges to the upper-bound
capacity defined for the uplink channel. More precisely,
\begin{equation}
\lim_{K\rightarrow{\infty}}{C_u(K)-R_{ICBS}(K)}=0,\label{eq_t61}
\end{equation}
where $C_u(K)=\frac{1}{2} \mathbb{E}_{\mathbf{H}} \left[
\max_{\begin{subarray}{c} \mathbf{Q}, \rm{Tr} \{\mathbf{Q}\} \leq P
\end{subarray}} \log \left(\left|\mathbf{I}_{KN}+\mathbf{H}
\mathbf{Q}\mathbf{H}^H\right| \right)\right]$ is the point to point
ergodic capacity of the uplink channel and $R_{ICBS}(K)$ is the
achievable rate of ICBS.
\end{thm}
\begin{proof}
By applying the cut-set bound theorem \cite{cover_book} on the
broadcast uplink channel, it can be easily verified
\cite{nabar},\cite{nabar2} that the point-to-point capacity of the
uplink channel, $C_u(K)$, is an upper-bound on the capacity of the
parallel MIMO relay network. Note that the factor $\frac{1}{2}$ in
the expression of $C_u(K)$ is due to the half-duplex relaying.
Define
$C_{u^{\star}}(K)=\frac{M}{2}\log\left(1+\frac{KNP}{M}\right)$. We
first show that $C_{u^{\star}}(K)$ is an upper-bound for $C_u(K)$,
and then prove that a lower-bound for $R_{ICBS}(K)$ converges to
$C_{u^{\star}}(K)$.
\begin{eqnarray}
C_u(K)&=&\frac{1}{2} \mathbb{E}_{\mathbf{H}} \left[
\max_{\begin{subarray}{c} \mathbf{Q} \\ \mbox{Tr} \{\mathbf{Q}\}
\leq P \end{subarray}}
\log \left(\left|\mathbf{I}_{KN}+\mathbf{H} \mathbf{Q}\mathbf{H}^H\right| \right)\right]\nonumber\\
&\stackrel{(a)}{=}&\frac{1}{2}\mathbb{E}_{\mathbf{H}}
\left[\max_{\begin{subarray}{c} \mathbf{Q} \\ \mbox{Tr}
\{\mathbf{Q}\}
\leq P \end{subarray}}\log \left(\left|\mathbf{I}_{M}+\mathbf{H}^H\mathbf{H} \mathbf{Q} \right| \right) \right]\nonumber\\
&\stackrel{(b)}{\leq}&\frac{1}{2}\mathbb{E}_{\mathbf{H}}
\left[\max_{\begin{subarray}{c} \mathbf{Q} \\ \mbox{Tr}
\{\mathbf{Q}\}
\leq P \end{subarray}}M\log\left(1+\frac{\mbox{Tr}\left\{{\mathbf{H}^H\mathbf{H} \mathbf{Q}}\right\}}{M} \right)\right]\nonumber\\
&\stackrel{(c)}{\leq}&\frac{M}{2} \mathbb{E}_{\mathbf{H}}
\left[\max_{\begin{subarray}{c} \mathbf{Q} \\ \mbox{Tr}
\{\mathbf{Q}\}
\leq P \end{subarray}}\log\left(1+\frac{\mbox{Tr}\left\{\mathbf{H}^H\mathbf{H}\right\}\mbox{Tr}\left\{\mathbf{Q}\right\}}{M} \right)\right]\nonumber\\
&\stackrel{(d)}{\leq}&\frac{M}{2}\log\left(1+\frac{P}{M}\mathbb{E}_{\mathbf{H}}\left[\mbox{Tr}\left\{\mathbf{H}^H\mathbf{H}\right\}\right] \right)\nonumber\\
&=&C_{u^{\star}}(K).\label{eq_t62}
\end{eqnarray}
Here, $(a)$ follows from the matrix determinant
equality\footnote{Assuming
 $\mathbf{A}$ and $\mathbf{B}$ to be $M\times{N}$ and $N\times{M}$ matrices respectively, we have $\left|\mathbf{I}_M+\mathbf{A}\mathbf{B}\right|=\left|\mathbf{I}_N+\mathbf{B}\mathbf{A}\right|$ \cite{matrix_book}.}
, $(b)$ results from the fact that for any positive semidefinite
matrix $\mathbf{A}$, we have $|\mathbf{A}|\leq\left(
\frac{\mbox{Tr}\left\{\mathbf{A}\right\}}{M}\right)^M$, $(c)$
follows from the generalization of the Cauchy-Schwarz inequality to
the positive semidefinite matrices\footnote{Assuming
 $\mathbf{A}$ and $\mathbf{B}$ to be positive semidefinite matrices respectively, we have
  $\mbox{Tr}\left\{\mathbf{A}\mathbf{B}\right\}\leq\mbox{Tr}\left\{\mathbf{A}\right\}\mbox{Tr}\left\{\mathbf{B}\right\}$
   \cite{convex_book}.}, and $(d)$ follows from the concavity of the logarithm function. Rephrasing
(\ref{eq:interfer}), we have
\begin{eqnarray}
\mathbf{y}&=&\alpha\mathbf{H}^{\star}\mathbf{x}^{\prime}+\mathbf{n}^{\star},
\end{eqnarray}
where
\begin{eqnarray}
\mathbf{H}^{\star}&=&\mathbf{\Lambda}^{\frac{1}{2}}
-\sum_{k\in\mathcal{A}}{\mathbf{U}_k^H\mathbf{H}_k\mathbf{V}},\\
\mathbf{n}^{\star}&=&\alpha
\sum_{k\in\mathcal{A}^c}{\mathbf{U}_k^H\mathbf{n}_k} + \mathbf{z}
\sim\mathcal{CN}\left(\mathbf{0},\mathbf{P}_{\mathbf{n}^{\star}}\right),
\end{eqnarray}
where
$\mathbf{P}_{\mathbf{n}^{\star}}=\alpha^2\left(\sum_{k\in\mathcal{A}^c}{\mathbf{U}_k^H\mathbf{U}_k}\right)+\mathbf{I}_M$.
The achievable rate of such a system is
\begin{eqnarray}
R_{ICBS}(K)&=&\frac{1}{2}\mathbb{E}_{\mathbf{H}}\left[\log\left(\left|\mathbf{I}_{M}+\alpha^2\frac{P}{M}\mathbf{H}^{\star}\mathbf{H}^{\star{H}}\mathbf{P}_{\mathbf{n}^{\star}}^{-1}\right|\right)\right]\nonumber\\
&\geq&\frac{1}{2}\mathbb{E}_{\mathbf{H}}\left[\log\left(\left|\alpha^2\frac{P}{M}\mathbf{H}^{\star}\mathbf{H}^{\star{H}}\mathbf{P}_{\mathbf{n}^{\star}}^{-1}\right|\right)\right]\nonumber\\
&\stackrel{(a)}{\geq}&\frac{1}{2}\mathbb{E}_{\mathbf{H}}\left[\log\left(\left|\frac{\alpha^2}{1+\alpha^2}\frac{P}{M}\mathbf{H}^{\star}\mathbf{H}^{\star{H}}\right|\right)\right]\nonumber\\
&=&\frac{M}{2}\log\left(\frac{\alpha^2}{1+\alpha^2}\right)+\frac{1}{2}\mathbb{E}_{\mathbf{H}}\left[\log\left(\left|\frac{P}{M}\mathbf{H}^{\star}\mathbf{H}^{\star{H}}\right|\right)\right],\label{eq:t6_appr1}
\end{eqnarray}
where $(a)$ follows from the fact that
$\mathbf{P}_{\mathbf{n}^{\star}}=(\alpha^2+1)\mathbf{I}_M-\alpha^2\left(\sum_{k\in\mathcal{A}}{\mathbf{U}_k^H\mathbf{U}_k}\right)$
which results in
$\mathbf{P}_{\mathbf{n}^{\star}}\preccurlyeq(\alpha^2+1)\mathbf{I}_M$,
or equivalently
$\mathbf{P}_{\mathbf{n}^{\star}}^{-1}\succcurlyeq\frac{1}{\alpha^2+1}\mathbf{I}_M$.
For convenience, let
$$R_{L}(K)=\frac{1}{2}\mathbb{E}_{\mathbf{H}}\left[\log\left(\left|\frac{P}{M}\mathbf{H}^{\star}\mathbf{H}^{\star{H}}\right|\right)\right].$$
Since $\alpha$ is lower-bounded by the inverse of the threshold as
$\alpha\geq\sqrt{\frac{P}{\beta}}$, we have
$\lim_{K\rightarrow\infty}{\frac{M}{2}\log\left(\frac{\alpha^2}{1+\alpha^2}\right)}=0$,
or equivalently
\begin{equation}
\lim_{K\rightarrow\infty}R_{ICBS}(K)-R_{L}(K)\geq0 .\label{eq_t63}
\end{equation}
Define the events $E_K$ and $F_K$ as
$E_K\equiv\left(\lambda_{\min}\left(\mathbf{H}\right)\gtrsim{KN}\left[1+O\left(\sqrt[4]{\frac{\log{K}}{K}}\right)\right]\right)$
and
$F_K\equiv\left(\left\|\mathbf{U}_{\mathcal{A}}^H\mathbf{H}_{\mathcal{A}}\right\|^2\leq{\frac{K}{\log^2(K)}}\right)$.
Consequently, we have
\begin{eqnarray} \label{eq: pekfk}
\mathbb{P}\left[E_K, F_K\right]&\stackrel{(a)}{\geq}&1-\mathbb{P}[{E}_K^c]-\mathbb{P}[{F}_K^c]\nonumber\\
&\stackrel{(b)}{\gtrsim}&1+O\left(\frac{1}{K\sqrt{\log{K}}}\right)+O\left(\frac{\log^4(K)}{\sqrt{K}}\right)\nonumber\\
&\thicksim&1+O\left(\frac{\log^4(K)}{\sqrt{K}}\right).
\end{eqnarray}
 Here, $(a)$ follows from union bound inequality and $(b)$ follows from
Lemmas 4 and 5. Assume the diagonal entries of $\mathbf{\Lambda}$
are ordered as
$\lambda_1(\mathbf{H})\geq\lambda_2(\mathbf{H})\geq\cdots\geq\lambda_M(\mathbf{H})$.
Thus, $R_L(K)$ can be lower bounded as

\begin{eqnarray}
R_{L}(K)&\geq&\frac{1}{2}\mathbb{P}\left[E_K,
F_K\right]\mathbb{E}_{\mathbf{H}}\left[\log\left(\left|\frac{P}{M}\mathbf{H}^{\star}\mathbf{H}^{\star{H}}\right|\right)\Bigg{|}E_K,
F_K\right]\nonumber\\
&=&\mathbb{P}\left[E_K,
F_K\right]\mathbb{E}_{\mathbf{H}}\left[\log\left(\left|\sqrt{\frac{P}{M}}\left(\mathbf{\Lambda}^{\frac{1}{2}}-\mathbf{U}_{\mathcal{A}}^H\mathbf{H}_{\mathcal{A}}\mathbf{V}\right)\right|\right)\Bigg{|}E_K,
F_K\right]\nonumber\\
&\stackrel{(a)}{\geq}&\mathbb{P}\left[E_K,
F_K\right]\mathbb{E}_{\mathbf{H}}\left[\log\left(\left(\frac{P}{M}\right)^{\frac{M}{2}}\left(\prod_{i=1}^{M}{\lambda_i^{\frac{1}{2}}\left(\mathbf{H}\right)}-{}\right.\right.\right.\nonumber\\
&&\left.\left.\left.{}-\sum_{i=1}^{M}{i!\binom{M}{i}\left\|\mathbf{U}_{\mathcal{A}}^H\mathbf{H}_{\mathcal{A}}\mathbf{V}\right\|_{\star}^{i}\prod_{j=1}^{M-i}{\lambda_j^{\frac{1}{2}}{\left(\mathbf{H}\right)}}}\right)\right)\Bigg{|}E_K,
F_K\right]\nonumber\\
&\stackrel{(b)}{\geq}&\mathbb{P}\left[E_K,
F_K\right]\mathbb{E}_{\mathbf{H}}\left[\log\left(\left(\frac{P}{M}\right)^{\frac{M}{2}}\prod_{i=1}^{M}{\lambda_i^{\frac{1}{2}}\left(\mathbf{H}\right)}\cdot{}\right.\right.\nonumber\\
&&\left.\left.{}\cdot\left(1-\sum_{i=1}^{M}{i!\binom{M}{i}\left(\frac{\left\|\mathbf{U}_{\mathcal{A}}^H\mathbf{H}_{\mathcal{A}}\right\|^2}{\lambda_{\min}\left(\mathbf{H}\right)}\right)^{\frac{i}{2}}}\right)\right)\Bigg{|}E_K,
F_K\right]\nonumber\\
&\stackrel{(c)}{\gtrsim}&\mathbb{P}\left[E_K,
F_K\right]\mathbb{E}_{\mathbf{H}}\left[\log\left(\left(\frac{P}{M}\right)^{\frac{M}{2}}\prod_{i=1}^{M}{\lambda_i^{\frac{1}{2}}\left(\mathbf{H}\right)}\cdot{}\right.\right.\nonumber\\
&&\left.\left.{}\cdot\left(1-\sum_{i=1}^{M}{i!\binom{M}{i}\left({N\log^2(K)\left[1+O\left(\sqrt[4]{\frac{\log{K}}{K}}\right)\right]}\right)^{\frac{-i}{2}}}\right)\right)\Bigg{|}E_K,F_K\right]\nonumber\\
&\gtrsim&\mathbb{P}\left[E_K,
F_K\right]\left\{\frac{M}{2}\log\left(\frac{P}{M}\right)
+\frac{1}{2}\sum_{i=1}^{M}{\mathbb{E}_{\mathbf{H}}
\left[\log{\left(\lambda_i\left(\mathbf{H}\right)\right)}\Big{|}E_K,F_K\right]}
\right.
- \nonumber\\
&&\left.{}  -\frac{M}{\sqrt{N} \log (K)} \left( 1+ O \left(
\frac{1}{\log (K)}\right)
\right) \right\} \label{eq:t6_appr2}\\
&\stackrel{(d)}{\gtrsim}&\mathbb{P}\left[E_K,
F_K\right]\Bigg\{\frac{M}{2}\log\left(\frac{P}{M}\right)+
\frac{M}{2}\log\left(KN \left[ 1+O\left(\sqrt[4]{\frac{\log{K}}{K}}\right)\right]\right)-{}\nonumber\\
&&\left.{}-\frac{M}{\sqrt{N} \log (K)} \left( 1+ O \left(
\frac{1}{\log (K)}\right)
\right) \right\} \label{eq:t6_appr3}\\
&\stackrel{(e)}{\gtrsim}&\left\{\frac{M}{2}\log\left(\frac{KNP}{M}\right)
+ O \left( \frac{1}{\log ( K)}\right) \right\}\mathbb{P}\left[E_K,
F_K\right]\nonumber\\
&\stackrel{(f)}{\gtrsim}&\left\{\frac{M}{2}\log\left(\frac{KNP}{M}\right)+O\left(\frac{1}{\log
( K)}\right) \right\} \left[ 1+ O \left( \frac{\log ^4 (K)}{\sqrt{K
}}\right)\right]\label{eq:t6_appr4}
\nonumber\\
&\stackrel{(g)}{\sim}&\frac{M}{2}\log\left(\frac{KNP}{M}\right) + O
\left( \frac{1}{\log (K)}\right).\label{eq_t64}
\end{eqnarray}
Here, $(a)$ follows from an upper-bound on the determinant expansion
\footnote{$\det{\left(A\right)}=\sum_{\pi}\left(-1\right)^{\sigma\left(\pi\right)}{a_{1\pi_1}a_{2\pi_2}{\cdots}a_{n\pi_n}}
\leq \sum_{\pi}{|a_{1\pi_1}a_{2\pi_2}{\cdots}a_{n\pi_n}|}$, where
$\sigma$ is the parity function of permutation.} of
$\mathbf{\Lambda}^{\frac{1}{2}}-\mathbf{U}_{\mathcal{A}}^H\mathbf{H}_{\mathcal{A}}\mathbf{V}$,
 expanded over all possible set entries between $\mathbf{\Lambda}$
and $\mathbf{U}_{\mathcal{A}}^H\mathbf{H}_{\mathcal{A}}\mathbf{V}$,
$(b)$ follows from the fact that the Frobenius norm of a matrix is
an upper-bound on the square of the maximum absolute value among its
entries and also $\forall{i}: \lambda_{i}
(\mathbf{H})\geq\lambda_{\min} (\mathbf{H})$, $(c)$ follows from the
fact that the expectation is derived conditioned on the events $E_K$
and $F_K$, $(d)$ holds due to the fact that conditioned on $E_K$, we
have
$\lambda_i\left(\mathbf{H}\right)\gtrsim{KN}\left[1+O\left(\sqrt[4]{\frac{\log{K}}{K}}\right)\right]$,
$(e)$ follows from the fact that $\log \left( 1+ O \left(
\sqrt[4]{\frac{\log (K)}{K}}\right)\right) \sim O \left(
\sqrt[4]{\frac{\log (K)}{K}}\right) \sim o \left( \frac{1}{\log^2
(K)}\right)$, $(f)$ results from (\ref{eq: pekfk}), and finally,
$(g)$ follows  from the fact that $O \left( \frac{\log ^4
(K)}{\sqrt{K }}\right) \sim o \left( \frac{1}{\log (K)}\right)$.
Now, defining
$R_{S}\left(K\right)=\frac{M}{2}\log\left(\frac{KNP}{M}\right)$,
according to  (\ref{eq_t63}) and (\ref{eq_t64}), we have
\begin{equation}
\lim_{K\rightarrow\infty}{R_{ICBS}(K)-R_S(K)}\geq{0}.\label{eq_t65}
\end{equation}
Furthermore, we have:
\begin{equation}
\lim_{K\rightarrow\infty}{C_{u^{\star}}(K)-R_S(K)}={0}.\label{eq_t66}
\end{equation}
Comparing (\ref{eq_t62}), (\ref{eq_t65}) and (\ref{eq_t66}), and
observing the fact that $C_{u}(K)\geq{C}_{ICBS}(K)$, results in
(\ref{eq_t61}) and this completes the proof.
\end{proof}
\begin{corr}
The capacity of parallel MIMO Relay network, the point-to-point
capacity of the cut-set defined on the uplink channel, the
achievable rate of amplify and forward relaying, and the achievable
rate of ICBS, all converge to
$\frac{M}{2}\log\left(\frac{KNP}{M}\right)$, as $K \to \infty$.
\end{corr}
\begin{proof}
Defining $C(K)$, $C_u(K)$, $R_{AF}(K)$, and $R_{ICBS}(K)$ as the
capacity of parallel MIMO Relay network, the point-to-point capacity
of the cut-set defined on the uplink channel, the achievable rate of
the amplify and forward relaying, and the achievable rate of ICBS,
respectively, it is clear that
\begin{equation}
R_{ICBS}(K){\leq}R_{AF}(K){\leq}C(K){\leq}C_u(K).
\end{equation}
Relying on Theorem \ref{thm:main_thm}, we know
\begin{equation}
\lim_{K\rightarrow{\infty}}{C_u(K)-R_S(K)}=\lim_{K\rightarrow{\infty}}{R_{ICBS}(K)-R_S(K)}=0.
\end{equation}
By observing that $R_{AF}(K)$ and $C(K)$ are sandwiched between
$R_{ICBS}(K)$ and $C_u(K)$, Sandwich theorem tells us that
\begin{equation}
\lim_{K\rightarrow{\infty}}{R_{AF}(K)-R_S(K)}=\lim_{K\rightarrow{\infty}}{C(K)-R_S(K)}=0.
\end{equation}
\end{proof}
\begin{corr}
Achievable rate of ICBS is at most
$O\left(\frac{1}{\log\left(K\right)}\right)$ below the upper-bound
corresponding to the cut-set defined on the point-to-point uplink
channel, i.e. $C_u(K)$.
\end{corr}
\begin{proof}
Following the proof of Theorem \ref{thm:main_thm}, we observe
\begin{equation} \label{eq: alireza3}
C_{u}(K)-R_{ICBS}(K)\leq{\Delta}R_1+{\Delta}R_2+{\Delta}R_3,
\end{equation}
where ${\Delta}R_1=\frac{M}{2}\log\left(1+\frac{1}{\alpha^2}\right)$
results from the approximation of the first term in
(\ref{eq:t6_appr1}), ${\Delta}R_2= O \left( \frac{1}{\log
(K)}\right)$ in (\ref{eq_t64}), and finally,
$\Delta{R_3}=\frac{M}{2}\log\left(1+\frac{M}{KNP}\right) \sim O
\left( \frac{1}{K}\right)$ is the difference between $C_{u^{*}}(K)$
and $R_S(K)$. We know that
$\alpha\geq\sqrt{\frac{P}{\beta}}=\sqrt{P\log(K)}$, and as a result,
${\Delta}R_1=\frac{M}{2}\log\left(1+\frac{1}{P\log(K)}\right) \sim O
\left( \frac{1}{\log (K)}\right)$. Comparing the values of
$\Delta{R_i}, 1\leq{i}\leq{3}$, we conclude that
$C_{u}(K)-R_{ICBS}(K)=O\left(\frac{1}{\log\left(K\right)}\right)$.
\end{proof}
Apart from increasing the rate, using parallel relays also increases
the reliability of the transmission. As the following corollary
shows, the probability of outage when sending information at the
rate $O\left(\frac{1}{\log\left(K\right)}\right)$ below the ergodic
capacity approaches zero, as $K \to \infty$.
\begin{corr}
Consider the parallel MIMO relay network and ICBS with the threshold
value $\beta=\frac{1}{\log\left(K\right)}$. We have
\begin{equation}
\mathbb{P}\left[\frac{1}{2}\log\left(\left|\mathbf{I}_{M}+\alpha^2\frac{P}{M}\mathbf{H}^
{\star}\mathbf{H}^{\star{H}}\mathbf{P}_{\mathbf{n}^{\star}}^{-1}\right|\right)\lesssim{}C_u(K)+
O\left(\frac{1}{\log\left(K\right)}\right)\right]\sim{}O\left(\frac{\log^4\left(K\right)}{\sqrt{K}}\right).\nonumber
\end{equation}
\end{corr}
\begin{proof}
Following the proof of Theorem 1, we observe this outage event is a
subset of $E_K^c \bigcup F_K^c$, whose probability is shown to be
$O\left(\frac{\log^4\left(K\right)}{\sqrt{K}}\right)$.
\end{proof}

 Another interesting result is that by increasing the number of
relays, each relay can operate with a much lower power as compared
to the transmitter, while the scheme achieves the optimum rate.
This shows another benefit of using many parallel relays in the
network.
\begin{thm}\label{thm:second_uneq_power}
Up to the point that
$P_r(K)=\omega\left(\frac{P}{K}\log^9\left(K\right)\right)$, the
achievable rate of ICBS satisfies
\begin{equation}
\lim_{K\rightarrow\infty}{R_{ICBS}(K)-C_u(K)}=
\lim_{K\rightarrow\infty}R_{ICBS}(K)-\frac{M}{2}\log\left(\frac{KNP}{M}\right)=0.
\end{equation}
\end{thm}
\begin{proof}
We use the same steps as the proof of Lemma 4 with the same values
of $\gamma$ and $\xi$. Rewriting (\ref{eq: alireza1}), we have
\begin{eqnarray}
\mathbb{P} [v > \xi] \leq MNK \log^2 (K) \left[\mathbb{P} [B_k]+
\frac{2 \log (K)}{K} \left(\mathbb{P} [B_k] + c_1 \sqrt{\delta} +
c_2 e^{-\frac{d}{\sqrt{\delta}}}\right)\right],
\end{eqnarray}
where $\delta = \frac{\gamma}{\beta}$. In order that the second term
in (\ref{eq:t6_appr4}) (or equivalently $\Delta R_2$ in (\ref{eq:
alireza3})) approaches zero, we must have $\mathbb{P} [E_K,F_K] \sim
1+o \left( \frac{1}{\log (K)}\right)$, which implies that
$\mathbb{P} [v
> \xi] \sim 1+o \left( \frac{1}{\log (K)} \right)$. From the above
equation, it follows that having $\beta \sim \omega \left(
\frac{\log^9 (K)}{K}\right)$ incurs that $\sqrt{\delta} =
\sqrt{\frac{\gamma}{\beta}} \sim o \left( \sqrt{\frac{\frac{2 \log
(K)}{K}}{\frac{\log^9 (K)}{K}}}\right)$, or equivalently,
$\sqrt{\delta} \log^3 (K) \sim o \left( \frac{1}{\log (K)} \right)$,
which results in $\mathbb{P} [v
> \xi] \sim 1+o \left( \frac{1}{\log (K)} \right)$. Moreover, the first term in (\ref{eq:t6_appr1})
(or equivalently $\Delta R_1$ in (\ref{eq: alireza3})) approaches
zero, if $P_r(K)=\omega(\beta)$ (or equivalently, $\alpha \sim
\omega (1)$). Therefore, having $P_r (K) \sim \omega \left(
\frac{\log^9 (K)}{K}\right) $, results in $\Delta R_1 , \Delta R_2
\to 0$, which implies that $\lim_{K \to \infty} C_{u} (K) - R_{ICBS}
(K) =0$.
\end{proof}
\begin{thm}
The proposed Cooperative Beamforming scheme and its variant achieve
the maximum multiplexing gain of the relay channel. More precisely:
\begin{equation}
\lim_{P\rightarrow\infty}{\frac{R_{CBS}(P)}{\log(P)}}=\frac{M}{2},
\end{equation}
and $\frac{M}{2}$ is the maximum achievable multiplexing gain of the
underlying half duplex system. (Here $R_{CBS}(P)$ is the achievable
rate of the proposed scheme for the given power constraint $P$.)
\end{thm}
\begin{proof}
We prove the theorem for CBS. The statements of the proof are also
valid for the variant of CBS. First of all, from the last theorem,
we have
\begin{eqnarray}
C_u(P)\leq{C_{u^{\star}}(P)}\stackrel{(a)}{\leq}\frac{M}{2}\log\left(\frac{2KNP}{M}\right)
=\frac{M}{2}\log\left(\frac{KN}{M}\right)+\frac{M}{2}\log(P) +
\frac{M}{2}.
\end{eqnarray}
Here, $(a)$ follows from the assumption that $P$ is large enough
such that we have $P\geq\frac{M}{KN}$. Thus, the maximum achievable
multiplexing gain is
\begin{equation}
r_{\max}=\lim_{P\rightarrow\infty}{\frac{C_u(P)}{\log(P)}}\leq\frac{M}{2}.\label{eq:m_gain_1}
\end{equation}
To prove the theorem, it is sufficient to show that the multiplexing
gain of CBS is lower bounded by $\frac{M}{2}$. To show this, we
lower-bound the achievable rate of the scheme as follows:
\begin{eqnarray} \label{eq: alireza2}
R_{CBS}(P)&=&\frac{1}{2}\mathbb{E}_{\mathbf{G},\mathbf{H}}\left[\log\left(\left|\mathbf{I}_M+\frac{\alpha^2}{1+\alpha^2}\frac{P}{M}\mathbf{\Lambda}\right|\right)\right]\nonumber\\
&\geq&\frac{1}{2}\mathbb{E}_{\mathbf{G},\mathbf{H}}\left[\log\left(\left|\frac{\alpha^2}{1+\alpha^2}\frac{P}{M}\mathbf{\Lambda}\right|\right)\right]\nonumber\\
&\geq&\frac{M}{2}\log(P)+\frac{M}{2}\mathbb{E}_{\mathbf{H}}\left[\log\left(\lambda_{\min}\left(\mathbf{H}\right)\right)\right]-\frac{M}{2}\log(M)-\frac{M}{2}\mathbb{E}_{\mathbf{G},\mathbf{H}}\left[\log\left(1+\frac{1}{\alpha^2}\right)\right]\nonumber\\
&\stackrel{(a)}{\geq}&\frac{M}{2}\log(P)+\frac{M^2}{2}\int_{x=0}^{1}e^{-x}\log(x)dx-\frac{M}{2}\log(M)-\frac{M}{2}\mathbb{E}_{\mathbf{G},\mathbf{H}}\left[\log\left(1+\frac{1}{\alpha^2}\right)\right]\nonumber\\
&\geq&\frac{M}{2}\log(P)-\frac{M^2}{2}-\frac{M}{2} \log
(M)-\frac{M}{2}\mathbb{E}_{\mathbf{G},\mathbf{H}}\left[\log\left(1+\frac{1}{\alpha^2}\right)\right],\label{eq:m_gain_2}
\end{eqnarray}
where $(a)$ follows from the fact that $\lambda_{\min} (\mathbf{H})
\geq \lambda_{\min} (\mathbf{W})$, where $\mathbf{W}$ is an
arbitrary $M \times M$ submatrix of $\mathbf{H}$, noting that
$f_{\lambda_{\min} (\mathbf{W})} (\lambda) = M e^{-M \lambda}$,
$\lambda > 0$.
 Now, defining
$x_{\alpha}=\mathbb{E}_{\mathbf{G},\mathbf{H}}\left[\log\left(1+\frac{1}{\alpha^2}\right)\right]$,
it is sufficient to show that $x_{\alpha}$ can be upper bounded by a
finite expression independent of $P$. Defining
$x_{\alpha,k}=\log{\left[1+\lambda_{\min}^{-1}\left(\mathbf{G}_k\right)\left(\left\|\mathbf{H}_k\right\|^2+\frac{1}{P}\right)\right]}$,
we have
\begin{eqnarray}
x_{\alpha}&=&\mathbb{E}_{\mathbf{G},\mathbf{H}}\left[\log\left(1+\frac{\max_{1\leq{k}\leq{K}}
{\mathbb{E}_{\mathbf{x}, \mathbf{n}_k}
\left[\left\|\mathbf{G}_k^{\dagger}\mathbf{U}_k^H\mathbf{r}_k\right\|^2
\right]}}{P}\right)\right]\nonumber\\
&=&\mathbb{E}_{\mathbf{G},\mathbf{H}}\left[\max_{1\leq{k}\leq{K}}{\log\left(1+\frac{\mathbb{E}_{\mathbf{x},
\mathbf{n}_k}
\left[\left\|\mathbf{G}_k^{\dagger}\mathbf{U}_k^H\mathbf{r}_k\right\|^2
\right]}{P}\right)}\right]\nonumber\\
&\stackrel{(a)}{\leq}&\mathbb{E}_{\mathbf{G},\mathbf{H}}\left[\max_{1\leq{k}\leq{K}}{\log\left(1+\lambda_{\min}^{-1}\left(\mathbf{G}_k\right)\left(\left\|\mathbf{H}_k\right\|^2+\frac{1}{P}\right)\right)}\right]\nonumber\\
&=&\mathbb{E}_{\mathbf{G},\mathbf{H}}\left[\max_{1\leq{k}\leq{K}}{x_{\alpha,k}}\right]\nonumber\\
&\stackrel{(b)}{\leq}&\mathbb{E}_{\mathbf{G},\mathbf{H}}\left[\sum_{k=1}^K{x_{\alpha,k}}\right]\nonumber\\
&=&K\mathbb{E}_{\mathbf{G},\mathbf{H}}\left[{x_{\alpha,k}}\right]\label{eq:m_gain_3}
\end{eqnarray}
Here, $(a)$ results from matrix product norm inequality and
independency of $\mathbf{n}_k$ from $\mathbf{H}_k$ and $\mathbf{x}$,
and $(b)$ follows from the fact that $x_{\alpha,k}$'s are
nonnegative i.i.d. random variables. Without loss of generality, we
can assume $P$ is large enough such that $P\geq{1}$. We can
upper-bound $\mathbb{E}\left[x_{\alpha,k}\right]$ as
\begin{eqnarray}
\mathbb{E}\left[x_{\alpha,k}\right]&=&\mathbb{E}_{\mathbf{G}_k,\mathbf{H}_k}\left\{\log{\left[1+\lambda_{\min}^{-1}\left(\mathbf{G}_k\right)\left(\left\|\mathbf{H}_k\right\|^2+\frac{1}{P}\right)\right]}\right\}\nonumber\\
&\stackrel{(a)}{\leq}&\mathbb{E}_{\mathbf{G}_k,\mathbf{H}_k}\left[\log{\left(1+\lambda_{\min}{\left(\mathbf{G}_k\right)}+\left\|\mathbf{H}_k\right\|^2\right)}\right]-\mathbb{E}_{\mathbf{G}_k}\left[\log{\left(\lambda_{\min}\left(\mathbf{G}_k\right)\right)}\right]\nonumber\\
&\stackrel{(b)}{\leq}&\mathbb{E}_{\mathbf{G}_k}\left[\lambda_{\min}{\left(\mathbf{G}_k\right)}\right]+\mathbb{E}_{\mathbf{H}_k}\left[\left\|\mathbf{H}_k\right\|^2\right]-\mathbb{E}_{\mathbf{G}_k}\left[\log{\left(\lambda_{\min}\left(\mathbf{G}_k\right)\right)}\right]\nonumber\\
&\stackrel{(c)}{\leq}&N+MN-M\int_{x=0}^{1}{e^{-Mx}\log(x)dx}\nonumber\\
&\leq&MN+M+N.\label{eq:m_gain_4}
\end{eqnarray}
Here, $(a)$ follows from the assumption that $P\geq{1}$, $(b)$
follows from the fact that $\log(1+x)\leq{x}$, and $(c)$ follows
from the fact that
$\mathbb{E}\left[\lambda_{\min}{\left(\mathbf{G}_k\right)}\right]
\leq \mathbb{E} \left[\frac{\|\mathbf{G}_k\|^2}{M} \right]=N$, and
also $(a)$ in (\ref{eq: alireza2}). Comparing (\ref{eq:m_gain_2}),
(\ref{eq:m_gain_3}), and (\ref{eq:m_gain_4}), we have
\begin{equation}
R_{CBS}\left(P\right)\geq\frac{M}{2}\log(P)+O(1).
\end{equation}
As a result
\begin{equation}
r_{CBS}=\lim_{P\rightarrow\infty}{\frac{C_u(P)}{\log(P)}}\geq\frac{M}{2}.\label{eq:m_gain_5}
\end{equation}
Comparing (\ref{eq:m_gain_1}) and (\ref{eq:m_gain_5}) completes the
proof.
\end{proof}
\textit{Remark -} It is claimed in \cite{nabar} that the proposed
BNOP scheme achieves the full multiplexing gain of $\frac{M}{2}$,
for $K \to \infty$. However, it should be mentioned that  this
result is not valid for the asymptotically large values of SNR, for
any fixed number of relays. Moreover, it can easily be shown that
the interference term increases linearly with SNR, and as a result,
the SINR term is limited by a constant value for large SNR values.
Therefore, the multiplexing gain of BNOP scheme is zero for any
fixed number of relays.
\section{Simulation Results}

\begin{figure}[hbt]
  \centering
  \includegraphics[scale=.5]{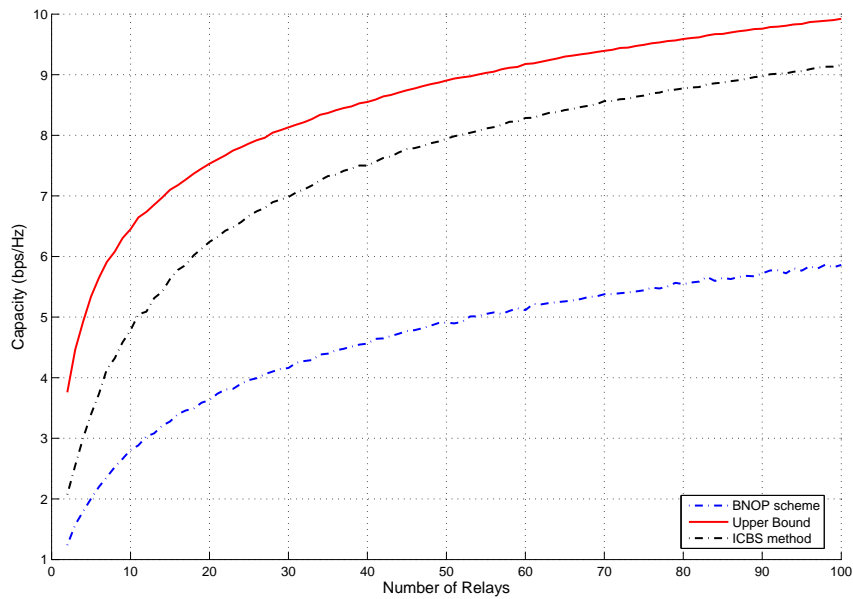}
\caption{Upper-bound of the capacity, ICBS, and BNOP matched
filtering Scheme vs. number of relays in parallel MIMO relay
network} \label{fig:num_result1}
\end{figure}

Figure \ref{fig:num_result1} shows the simulation results for the
achievable rate of ICBS, BNOP matched filtering scheme \cite{nabar},
and the upper-bound of the capacity based on the uplink Cut-Set for
varying number of relays. The number of transmitting and receiving
antennas in the relays, the transmitter, and the receiver is
$M=N=2$, and the SNR is $P_s=P_r=10dB$. While both of the schemes
demonstrate logarithmic scaling of rate in terms of $K$, we observe
that there is a significant gap between the BNOP scheme and our
scheme, reflecting the gap of $O(1)$ in the achievable rate of
\cite{nabar}. On the other hand, the gap between ICBS and the
upper-bound rapidly approaches zero due to the term
$O\left(\frac{1}{\log\left(K\right)}\right)$ predicted in Corollary
2.

\section{Conclusion}
A simple new scheme, Cooperative Beamforming Scheme (CBS), based on
Amplify and Forward (AF) strategy is introduced in a parallel MIMO
relay network. A variant of CBS, called Incremental Cooperative
Beamforming Scheme (ICBS) is shown to achieve the capacity of
parallel MIMO relay network for $K \to \infty$. The scheme is shown
to rapidly approach the upper-bound of the capacity with a gap no
more than $O\left(\frac{1}{\log\left(K\right)}\right)$. As a result,
it is shown that the capacity of a parallel MIMO relay network is
$C(K)=\frac{M}{2}\log\left(1+\frac{KNP}{M}\right)+O\left(\frac{1}{\log\left(K\right)}\right)$
in terms of the number of relays, $K$. Moreover, it is shown that as
the number of relays increases, the relays in ICBS can operate using
much less power without any performance degradation. Finally, the
proposed scheme is shown to achieve the maximum multiplexing gain
regardless of the number of relays. The simulation results confirm
the validity of the theoretical arguments.

\section*{Appendix A}\label{ap:a}
\subsection*{Proof of Lemma 2}
Let us denote $\mathbf{W}_i$ as the $i$th column of $\mathbf{W}$. In
\cite{alireza}, it has been shown that
\begin{eqnarray} \label{fw}
f_{\|\mathbf{W}_i\|^2} (x)= \frac{\Gamma (NK)}{\Gamma (N) \Gamma
(NK-N)} x^{N-1} (1-x)^{NK-N-1}, \quad i=1, \cdots, M,
\end{eqnarray}
which corresponds to the Beta distribution with parameters $N$ and
$NK-N$. Therefore, we have
\begin{eqnarray}
\mathbb{P} \left [ \|\mathbf{W}\|^2 \geq \gamma \right ] &=&
\mathbb{P}
\left [ \sum_{i=1}^M \|\mathbf{W}_i\|^2 \geq \gamma \right ]  \notag\\
&\leq& \mathbb{P} \left [ \max_{i} \|\mathbf{W}_i\|^2  \geq
\frac{\gamma}{M} \right ]  \notag\\
&=& \mathbb{P} \left [ \bigcup_{i=1}^M \mathcal{F}_i\right
] \notag\\
&\stackrel{(a)}{\leq}& M \mathbb{P} \left [ \mathcal{F}_i \right ],
\end{eqnarray}
where $(a)$ results from the Union bound on the probability, and
$\mathcal{F}_i \equiv \|\mathbf{W}_i\|^2 \geq \frac{\gamma}{M}$.
Defining $\gamma' \triangleq \frac{\gamma}{M}$, and using
(\ref{fw}), we obtain
\begin{eqnarray}
\mathbb{P} \left [ \|\mathbf{W}\|^2 \geq \gamma \right ] &\leq& M
\left( 1-F_{\|\mathbf{W}_i\|^2}
(\gamma')\right) \notag\\
&=& M \frac{\Gamma (NK)}{\Gamma (N) \Gamma (NK-N)}
\int_{\gamma'}^1 x^{N-1} (1-x)^{NK-N-1} dx \notag\\
&\stackrel{(a)}{=}& M \frac{\Gamma (NK)}{\Gamma (N) \Gamma (NK-N)}
\left(
\frac{\gamma'^{(N-1)} (1-\gamma')^{NK-N}}{NK-N} \right. + \notag\\
&& \left. \frac{1}{NK-N} \sum_{n=1}^{N-1} \left[\prod_{j=1}^{n}
\frac{(N-j)}{(NK-N+j)} \right]\gamma'^{(N-n-1)}
(1-\gamma')^{NK-N+n}\right) \notag\\
&=& M \sum_{n=1}^{N} \frac{(NK-1)!}{(N-n)! (NK-N+n-1)!}
\gamma'^{N-n} (1-\gamma')^{NK-N+n-1} \notag\\
&\leq& M \sum_{n=1}^{N} \frac{(NK \gamma')^{N-n}
(1-\gamma)^{NK-N}}{(N-n)!} \notag\\
 &\stackrel{(b)}{\sim}& \frac{M (NK
\gamma')^{N-1} (1-\gamma')^{NK-N}}{(N-1)!}
\left[ 1+O \left( \frac{1}{K \gamma'}\right)\right] \notag\\
&\sim& O \left( (K \gamma)^{N-1} e^{-\frac{\gamma}{M}NK} \right),
\end{eqnarray}
where $(a)$ follows from the integration by part, and $(b)$ follows
from the fact that $K \gamma' \sim \omega (1)$.
\rightline{$\blacksquare$}
\section*{Appendix B}\label{ap:b}
\subsection*{Proof of Lemma 5} The $(i,j)$th entry of $\mathbf{A}
\mathbf{A}^H$, denoted as $[\mathbf{AA}^H]_{i,j}$, can be written as
\begin{eqnarray}
[\mathbf{AA}^H]_{i,j} = \mathbf{a}_i \mathbf{a}_j^H,
\end{eqnarray}
where $\mathbf{a}_i$ is the vector representing the $i$th row of
$\mathbf{AA}^H$. Let us define $\mathbf{B}$ as
\begin{eqnarray}
\mathbf{B} \triangleq [\mathbf{b}_1^T| \cdots | \mathbf{b}_r^T]^T,
\end{eqnarray}
where $\mathbf{b}_i = \dfrac{\mathbf{a}_i}{\|\mathbf{a}_i\|}$, $i=1,
\cdots, r$. We have
\begin{eqnarray}
[\mathbf{BB}^H]_{i,j} = \left [ \begin{tabular} {lc} $1$ & $i=j$
\\
$\gamma(i,j)$& $i \neq j$
\end{tabular} \right. ,
\end{eqnarray}
where $\gamma(i,j) \triangleq \mathbf{b}_i \mathbf{b}_j^H =
\dfrac{\mathbf{a}_i \mathbf{a}_j^H
}{\|\mathbf{a}_i\|\|\mathbf{a}_j\|}$. The pdf of $z(i,j)=|\gamma
(i,j)|^2$ has been computed in \cite{alireza}, Lemma 3, as
\begin{eqnarray} \label{zij}
p_{z(i,j)} (z) = (s-1) (1-z)^{s-2}.
\end{eqnarray}
Let us define $\mathcal{C}$ as the event that $z(i,j) <
\frac{1}{\sqrt{s}}$ for all $i \neq j$. Using (\ref{zij}), we have
\begin{eqnarray} \label{pc}
\mathbb{P} [\mathcal{C}] &=& \mathbb{P} \left [ \bigcap_{i \neq j}
\left(z(i,j) <
\frac{1}{\sqrt{s}}\right) \right ]  \notag\\
&\stackrel{(a)}{\geq}&
1-\frac{r(r-1)}{2}\left(1-\frac{1}{\sqrt{s}}\right)^{s-1} \notag\\
&\sim& 1+O (e^{-\sqrt{s}}),
\end{eqnarray}
where $(a)$ results from the Union bound on the probability, noting
that $z(i,j)=z(j,i)$, $\forall i,j$. Conditioned on $\mathcal{C}$,
the orthogonality defect of $\mathbf{B}$, defined as
$\frac{\prod_{i=1}^r \|\mathbf{b}_i\|^2}{\left|\mathbf{BB}^H
\right|}$, can be written as
\begin{eqnarray}
\delta_{\mathcal{C}} (\mathbf{B}) &=&
\frac{1}{\left|\mathbf{BB}^H \right|} \notag\\
&=& \frac{1}{1+O (\frac{1}{\sqrt{s}})} \notag\\
&=& 1+O \Big( \frac{1}{\sqrt{s}} \Big),
\end{eqnarray}
where $\delta_{\mathcal{C}} (\mathbf{B})$ denotes the orthogonality
defect of $\mathbf{B}$, conditioned on $\mathcal{C}$. Hence, using
the fact that the orthogonality defect of $\mathbf{A}$ and
$\mathbf{B}$ are equal, conditioned on $\mathcal{C}$ we can write
\begin{eqnarray} \label{prod}
\prod_{i=1}^{r} \lambda_i (\mathbf{A})&=& \left|\mathbf{AA}^H \right| \notag\\
&=& \prod_{i=1}^{r} \|\mathbf{a}_i\|^2 \left[ 1+O\Big(
\frac{1}{\sqrt{s}}\Big)\right],
\end{eqnarray}
where $\lambda_i (\mathbf{A})$'s denote the singular values of
$\mathbf{AA}^H$. Moreover,
\begin{eqnarray} \label{sum}
\sum_{i=1}^r \lambda_i (\mathbf{A}) &=& \mbox{Tr} \{\mathbf{AA}^H\}
\notag\\
&=& \sum_{i=1}^{r} \|\mathbf{a}_i\|^2.
\end{eqnarray}
Now, let us define events $\mathcal{D}_i$ as follows:
\begin{eqnarray}
\mathcal{D}_i \equiv  \left\{s (1-\epsilon) < \|\mathbf{a}_i\|^2 < s
(1+\epsilon) \right\}, \quad i=1, \cdots, r,
\end{eqnarray}
where $\epsilon \triangleq \sqrt{\frac{2 \log (s)}{s}}$. Since
$\|\mathbf{a}_i\|=\sum_{j=1}^s |a_{i,j}|^2$, where $a_{i,j}$ denotes
the $(i,j)$th entry of $\mathbf{A}$, and having the fact that
$|a_{i,j}|^2$ are i.i.d. random variables with unit mean and unit
variance, using Central Limit Theorem (CLT), $\frac{1}{s}
\|\mathbf{a}_i\|^2$ approaches, in probability, to a Gaussian
distribution with unit mean and variance $\frac{1}{s}$, as $s$ tends
to infinity. More precisely, defining $X \triangleq
\frac{\frac{1}{s}\|\mathbf{a}_i\|^2}{\sqrt{\frac{1}{s}}}$ and using
Theorem 5.24 in \cite{clt}, we have
\begin{eqnarray}
\mathbb{P} \left[ -\sqrt{2 \log (s)} < X < \sqrt{2 \log (s)} \right]
&=& 1- \left[ 1- \Phi \left(\sqrt{2 \log (s)}\right)\right] \exp
\left\{ \frac{\gamma_3  \sqrt{2} \sqrt{\log^3 (s)}}{3 \sigma^3
\sqrt{s}}\right\} - \nonumber\\
&& {}- \Phi \left(-\sqrt{2 \log (s)}\right) \exp \left\{-
\frac{\gamma_3  \sqrt{2} \sqrt{\log^3 (s)}}{3 \sigma^3
\sqrt{s}}\right\} + \nonumber\\
&& {} + O \left( s^{-1/2} e^{- \log (s)}\right) \notag\\
&\stackrel{(a)}{\approx}& 1-\frac{1}{s \sqrt{ \pi \log (s)}  }
\left[ 1+O \left( \sqrt{\frac{\log^3 (s)}{s}}\right)\right] + O
\left( \frac{1}{s \sqrt{s}}\right), \nonumber
\\
\end{eqnarray}
where $\Phi (.)$ denotes the CDF of the normal distribution, and
$\sigma^2$ and $\gamma_3$ denote the second and third moments of
$|a_{i,j}|^2$, respectively. $(a)$ follows from the approximation of
$\Phi (x)$ for large $x$ by $1-\frac{1}{\sqrt{2 \pi} x}
e^{-\frac{x^2}{2}}$ and the fact that $\sigma \sim \gamma_3 \sim
\Theta (1)$. From the above equation, $\mathbb{P} [\mathcal{D}_i]$
can be computed as
\begin{eqnarray} \label{pd}
\mathbb{P} [ \mathcal{D}_i ] &=& \mathbb{P} \left [ 1-\epsilon <
\frac{1}{s} \|\mathbf{a}_i\|^2 < 1+\epsilon \right ] \notag\\
&\sim& 1+ O \left(\frac{1}{s \sqrt{ \log (s)}} \right),
\end{eqnarray}
in which we have used the definition of $\epsilon$ which is
$\sqrt{\frac{2 \log (s)}{s}}$. Conditioned on $\mathcal{C}$ and
$\mathcal{D}$, where $\mathcal{D} \triangleq \bigcap_{i=1}^r
\mathcal{D}_i$, and using (\ref{prod}) and (\ref{sum}), we can write
\begin{eqnarray} \label{eta}
\eta &\triangleq& \frac{\prod_{i=1}^r
\lambda_i}{\overline{\lambda}^r} \notag\\
&=& \frac{\prod_{i=1}^r \left[s (1+O(\epsilon)) \right] \left[1+O
\Big( \frac{1}{\sqrt{s}} \Big) \right]}{\left[\frac{1}{r}
\sum_{i=1}^r s
(1+O(\epsilon))\right]^r} \notag\\
&=& 1+O(\epsilon) \notag\\
&=& 1+O \Big( \sqrt{\frac{\log (s)}{s}}\Big),
\end{eqnarray}
where $\overline{\lambda} \triangleq \frac{1}{r} \sum_{i=1}^r
\lambda_i$. Suppose that $\lambda_{\min} = \alpha
\overline{\lambda}$ ($\alpha < 1$). We have
\begin{eqnarray}
\eta &\stackrel{(a)}{\leq}& \frac{\alpha \overline{\lambda}
\left[\frac{1}{r-1} (r \overline{\lambda} - \alpha
\overline{\lambda})\right]^{r-1} }{\overline{\lambda}^r} \notag\\
&=& \frac{\alpha (r-\alpha)^{r-1}}{(r-1)^{r-1}},
\end{eqnarray}
where $(a)$ follows from the fact that knowing $\lambda_{\min}$, the
product of the rest of the singular values is maximized when they
are all equal. Hence, having the sum constraint of $r
\overline{\lambda}$ yields $\prod_{i=1}^r \lambda_i < \alpha
\overline{\lambda} \left[\frac{1}{r-1} (r \overline{\lambda} -
\alpha \overline{\lambda})\right]^{r-1} $. Using (\ref{eta}), and
noting that $f(\alpha) \triangleq \frac{\alpha
(r-\alpha)^{r-1}}{(r-1)^{r-1}}$ is an increasing function of
$\alpha$ over the interval $[0,1]$, and writing the Taylor series of
$f(\alpha)$ about 1, noting $f' (1)=0$ and $f''(1)=\frac{-r}{r-1}$,
we have
\begin{eqnarray}
\frac{\alpha (r-\alpha)^{r-1}}{(r-1)^{r-1}} &=& 1+O \Bigg(
\sqrt{\frac{\log (s)}{s}}\Bigg). \notag\\
\Rightarrow \frac{r (1-\alpha)^2}{2(r-1)} &\sim& O \Bigg(
\sqrt{\frac{\log (s)}{s}}\Bigg). \notag\\
\Rightarrow  \alpha &\sim& 1+O \Bigg( \sqrt[4]{\frac{\log
(s)}{s}}\Bigg).
\end{eqnarray}
In other words, conditioned on $\mathcal{C}$ and $\mathcal{D}$, it
follows that $\lambda_{\min} = \overline{\lambda} \left[ 1+O \Big(
\sqrt[4]{\frac{\log (s)}{s}}\Big)\right]$. Moreover, conditioned on
$\mathcal{D}$, we have $\overline{\lambda} = s \left[ 1+O \Big(
\sqrt{\frac{\log (s)}{s}}\Big)\right]$. As a result,
\begin{eqnarray}
\mathbb{P} \left [ \lambda_{\min} \sim s \left[ 1+O \Bigg(
\sqrt[4]{\frac{\log (s)}{s}}\Bigg)\right]\right ] &\geq& \mathbb{P}
[\mathcal{C}
\cap \mathcal{D}] \notag\\
&\stackrel{(a)}{=}& \mathbb{P} [\mathcal{C}] \mathbb{P} [\mathcal{D}] \notag\\
&\stackrel{(b)}{=}& \mathbb{P} [\mathcal{C}] \left(\mathbb{P}
[\mathcal{D}_i] \right)^r \notag\\
&\stackrel{(\ref{pc}), (\ref{pd})}{\sim}& \left[
1+O(e^{-\sqrt{s}})\right] \left[ 1+ O \left(\frac{1}{s \sqrt{ \log
(s)}} \right)
\right]^r \notag\\
&\sim& 1+O \Big(\frac{1}{s \sqrt{\log (s)}} \Big),
\end{eqnarray}
where $(a)$ follows from the fact that the norm and direction of a
Gaussian vector are independent of each other, and as a result,
$\mathcal{C}$ and $\mathcal{D}$ are independent. $(b)$ follows from
the fact that $\mathcal{D}_i$'s are independent and have the same
probability. \\
\rightline{$\blacksquare$}

\bibliographystyle{IEEEtran}

\end{document}